\documentclass[aps,amsmath,amssymb,superscriptaddress,twocolumn,showpacs,12pts, floatfix, showpacs,intlimits]{revtex4-1}

\usepackage[english]{babel}
\usepackage[utf8]{inputenc}
\usepackage{float}
\usepackage{amsthm}
\usepackage{mathtools}
\usepackage{physics}
\usepackage{xcolor}
\usepackage{graphicx}
\usepackage[left=23mm,right=13mm,top=35mm,columnsep=15pt]{geometry} 
\usepackage{adjustbox}
\usepackage{placeins}
\usepackage[T1]{fontenc}
\usepackage{lipsum}
\usepackage{csquotes}
\begin{document}
\title{Quantum-inspired protocol for measuring the degree of similarity between spatial shapes}

\author{Daniel F. Urrego,}
\email[Correspondence email address: ]{daniel.urrego@icfo.eu}% Your name
\affiliation{ICFO -- Institut de Ciencies Fotoniques, The Barcelona Institute of Science and Technology, 08860 Castelldefels, Barcelona, Spain.\looseness=-1}
\author{Juan P. Torres}
\affiliation{ICFO -- Institut de Ciencies Fotoniques, The Barcelona Institute of Science and Technology, 08860 Castelldefels, Barcelona, Spain.\looseness=-1}
\affiliation{Department of Signal Theory and Communications, Universitat Politecnica de Catalunya, 08034 Barcelona, Spain\looseness=-1}

\date{\today} % Leave empty to omit a date

\begin{abstract}
We put forward and demonstrate experimentally a {\it quantum-inspired} protocol that allows to quantify the degree of similarity between two spatial shapes embedded in two optical beams without the need to measure the amplitude and phase across each beam. Instead the sought-after information can be retrieved measuring the degree of polarization of the combined optical beam, a measurement that is much easier to implement experimentally. The protocol makes use of non-separable optical beams, whose main trait is that different degrees of freedom (polarization and spatial shape here) can not be described independently.One important characteristic of the method described is that it allows to compare two unknown spatial shapes.
\end{abstract}

%\keywords{first keyword, second keyword, third keyword}

\maketitle

\section{Introduction}
Entanglement is a genuine quantum feature that describes the quantum correlations that exist between physically separated systems whose origin cannot be explained with classical physics concepts. E. Schrodinger~\cite{Schrodinger1935} defined entanglement as ``the characteristic trait of quantum mechanics, the one that enforces its entire departure from classical lines of thought".  For the sake of simplicity, let us consider a pure state $|\Psi \rangle$ composed of two subsystems, $A$ and $B$, with finite dimensions $d$. Entanglement between subsystems $A$ and $B$ is related to the idea of separability. One can always write the quantum state as a Schmidt decomposition  \cite{peres1995}
\begin{equation}
    |\Psi\rangle_{AB}=\sum_{i=1}^{d}  \sqrt{\lambda_i} |u_i\rangle_A   |v_i\rangle_B, 
    \label{quantum}
\end{equation}
where $|u_i\rangle_A$ and $|v_i\rangle_B$ constitute  orthonormal basis in the Hilbert spaces of subsystems  $A$ and $B$, and $\lambda_i$ are real and positive coefficients normalized so that $\sum_i \lambda_i=1$. If the Schmidt decomposition contains a single term, i.e., $\lambda_1=1$ and $\lambda_i=0$ for $i \ne 1$, the state is said to be separable, or not entangled. Otherwise the quantum state is entangled. A maximally entangled state corresponds to the case $\lambda_i=1/d$ for $i=1\dots d$.

The idea of separability also plays an important role in  classical optics~\cite{Spreeuw1998}, although now we consider different degrees of freedom of an individual system. One can generate optical beams where the different degrees of freedom that describe the beam cannot be considered separately \cite{McLaren2015,perumangatt2015,salla2015,Zhao2020}. For instance, considering the polarization and spatial shape degrees of freedom, one can generate \cite{Valles2014,McLaren2015,leuchs2016,Ndagano2016,eberly2017} an optical beam with electric field $E({\bf r})$,
\begin{equation}
    \vec{E}({\bf r})= \Big[ \sqrt{I_H}\, \Psi_H({\bf r})\, \hat{H}+ \sqrt{I_V}\, \Psi_V({\bf r})\, \hat{V} \Big] \exp (ikz),
     \label{classical}
\end{equation}
where ${\bf r}=(x,y)$ is the transverse position, $z$ is the direction  of propagation, $k$ is the wavenumber and ${\hat H}$ and ${\hat V}$ refer to horizontal and perpendicular polarizations. $I_{H,V}$ are the intensities of the two orthogonal polarizations, and $\Psi_H({\bf r})$ and $\Psi_V({\bf r})$ are the normalized spatial shapes of each polarization component. 

The analysis of the similarities and differences between Eqs.~(\ref{quantum}) and (\ref{classical}), and its implications and interpretations, has driven intense research in the last two decades \cite{Spreeuw2001,Kaltenbaek2009,Simon2010,Qian2011,Zela2014,Kagalwala2013,Svozilik2015}. Some researchers refer to the non-separability of Eq.~(\ref{classical}) as {\it nonquantum} or {\it classical} entanglement \cite{Qian2015, Qian2017}, although others claim that the use of the word entanglement in a non-quantum context might be misleading \cite{Karimi2015}. Non-separable optical beams have been tested using Bell's inequality experimental scenarios, obtaining similar results to its quantum counterpart~\cite{Borges2010,Kagalwala2013,chowdhury2013,Valles2014,Stoklasa2015,chithrabhanu2016,prabhakar2015}. However, the interpretation of these results is different when using single-photon detections and field intensity measurements \cite{Markiewicz2019}, leaving no place for the idea of {\it violation} of Bell's inequality in the classical case. How far one can go in the analogy between the quantum and classical scenarios is a matter of discussion and controversy in the scientific community.

The analogies and differences between classical and quantum separability allows one to bring questions and solutions from the quantum domain into the classical one \cite{Kwiat2000,Collins2002,Grover2002,Bhattacharya2002,Erkmen2006,Banaszek2007}. For instance, by inspecting how quantum principal component analysis achieves exponential speedup, Tang \cite{tang2021} describes an analog classical protocol that is only polynomially slower than its quantum counterpart. Leaving out the implications for interpretations and determining what is {\it uniquely} quantum mechanical, quantum ideas can serve as a source of inspiration for new classical solutions, the so-called {\it quantum-inspired} protocols \cite{Kaltenbaek2008,Lavoie2009,Mazurek2013,Ogawa2015,Altmann2018, Urrego2020}. These classical protocols based on non-separable optical beams have been used to encode information in optical communications~\cite{Milione2015, Milione2015a}, to characterize quantum channels \cite{Ndagano2017}, to perform state transfer~\cite{Hashemi2015}, to do stimulated emission depletion (STED) microscopy~\cite{Torok2004} and for sensing and metrology~\cite{Toppel2014,Berg-Johansen2015}.

Here we demonstrate a quantum inspired protocol that aims at measuring the degree of similarity between two spatial shapes without measuring the beam profiles. The direct measurement of the profile of a light beam is always a cumbersome measurement since it implies measuring amplitudes and phases across the whole beam. Instead we obtain the sought-after information by measuring only the degree of polarization  of an optical beam, a measurement easily done with the help of half-wave plates and polarizing beam splitters. The protocol considered is inspired by the analysis of the first-order coherence of subsystems of a bipartite quantum entangled state. We will see that the protocol proposed here is an example where classical physics seems to mimic certain aspects of a {\it truly} quantum effect such it is entanglement. 

The method proposed described below can be used for pattern recognition, where an input image (spatial shape) is compared with a set of images (spatial shapes) contained in a database. The key enabling factor would be to embed the input image, and each of the images in the database, with orthogonal polarizations, measuring the global polarization state afterwards as it will be described below. It is well-known that alternative techniques, such as the use of matched filters \cite{goodman2005} can also do pattern recognition. Both techniques share a technical hurdle for its practical implementation: large databases would require the implementation of correspondingly large databases of spatial shapes.

However, the method proposed here shows a fundamental, and important, advantage that alternative methods do not have. It allows to determine the degree of similarity of two unknown spatial shapes. This is a characteristic that the classical and quantum scenarios share. In the quantum scenario, we retrieve information about two unknown polarization states of photon $B$ by measuring the polarization state of photon $A$. In the classical scenario, we retrieve information about two unknown spatial shapes, measuring the global state of polarization of the combined optical beam.

\section{Description of the protocol for measuring the degree of similarity between two spatial shapes by measuring the degree of polarization}
The protocol we put forward here originates from the following observation. When detecting the intensity of the entire non-uniform optical beam, its state of polarization can be described by the coherence matrix \cite{mandel1995,Zela2014}
\begin{equation}
   J=\left( \begin{tabular}{cc}
        $I_H$ &  $ \sqrt{I_H I_V}\, \gamma$    \\
         $\sqrt{I_H I_V}\,\gamma^*$ &  $I_V $ 
   \end{tabular}\right), 
   \label{coherence_matrix}
\end{equation}
where $\gamma=\int d\textbf{r}\,\Psi_{H}({\bf r})\Psi^{*}_{V}({\bf r})$ is the overlap between the spatial shapes $\Psi_H({\bf r})$ and $\Psi_V({\bf r})$ of the horizontal and vertical components. The condition for a completely polarized wave is \cite{mandel1995}  $\text{det}~J=I_H I_V\, (1-|\gamma|^2)=0$ which implies that the two spatial shapes should be equal. In general, the degree of polarization $P$ of the beam is
\begin{equation}
    P=\Big[ 1-\frac{4\,\text{det}~J}{(\text{Tr} ~J)^2}\Big]^{1/2}=\Big[ 1-\frac{4I_H I_V\,(1-|\gamma|^2)}{(I_H+I_V)^2}\Big]^{1/2}.
    \label{P}
\end{equation}
If we can measure the degree of polarization and determine the overlap $\gamma$ we can determine the degree of similarity between two spatial shapes without measuring the amplitude and phase at each location of the two spatial shapes. When $I_H=I_V$, the degree of polarization is equal to the overlap, i.e., $P=|\gamma|$.

This analysis is reminiscent of what happens with a two-photon quantum entangled state in polarization of the form
\begin{equation}
    |\Psi\rangle_{AB}=\alpha |w_1\rangle_A   |H\rangle_B+\beta  |w_2\rangle_A   |V\rangle_B,   
\end{equation}
where $|w_1 \rangle$ and $|w_2 \rangle$ designate two polarization states of photon $A$, in general non-orthogonal, and the coefficients $\alpha$ and $\beta$ are the weights of the two components of the state $|\Psi\rangle_{AB}$ with $|\alpha|^2+|\beta|^2=1$. The quantum state of photon $B$ is
\begin{equation}
   \rho_B=\left( \begin{tabular}{cc}
        $|\alpha|^2$ &  $ \alpha^* \beta \,\langle w_1| w_2 \rangle$    \\
         $\alpha \beta^* \,\langle w_2| w_1 \rangle$ &  $|\beta|^2$ 
   \end{tabular}\right), 
   \label{density_matrix}
\end{equation}
Notice the remarkable similarity between Eqs. (\ref{coherence_matrix}) and (\ref{density_matrix}). The equivalent expression of the the degree of polarization given in Eq. (\ref{P}) is the purity $p$ of the state
\begin{equation}
    p=1-|\alpha|^2||\beta|^2 (1-|\Gamma|^2), 
\end{equation}
where we have defined  $\Gamma=\langle w_1|w_2\rangle$. Notice that the parameters $\gamma$ (classical scenario) and $\Gamma$ (quantum scenario) play similar roles. For arbitrary single-photon states, the purity of the quantum state is a measure of the degree of first-order coherence of the state \cite{glauber1966}.

\begin{figure}[t!]
\centering
\includegraphics[scale=0.5]{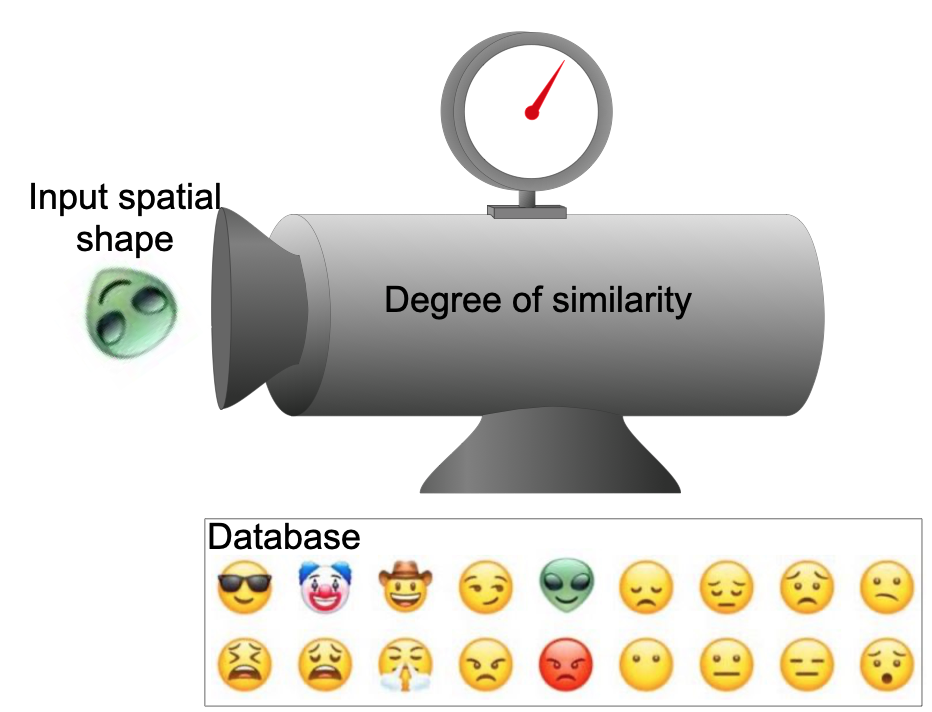}
\caption{Sketch of the protocol. We want to determine is an input spatial shape has an equal counterpart in a database that contains many spatial shapes for comparison. In a more general scenario, we want to quantify the degree of similarity of the input shape with all of the spatial shapes contained in the database.}
\label{fig: cartoon}
\end{figure}
Let us describe now the theoretical basis of the protocol we put forward here.  We consider that we need to determine if an input spatial shape has an equal counterpart in a database that contains a myriad of spatial shapes (see Fig.~\ref{fig: cartoon}). The input spatial shape has to be compared with each one of the spatial shapes contained in the database. The input spatial shape, given by $\Psi_H({\bf r})$, is embedded in an optical beam with horizontal polarization. A spatial shape of the database, given by $\Psi_V({\bf r})$, is embedded in an optical beam with vertical polarization. The two light beams enter a polarizing beam splitter (PBS) that projects into diagonal $\hat{D}=\frac{1}{\sqrt{2}}(\hat{H}+\hat{V})$ and antidiagonal $\hat{A}=\frac{1}{\sqrt{2}}(\hat{H}-\hat{V})$ polarizations. The total intensity detected by a bucket detector is
\begin{equation}
    I(\phi)=\frac{1}{2}\Big(I_H+I_V+2\sqrt{I_H I_V} |\gamma|\cos(\phi+\theta)\Big),
\end{equation}
where $\gamma=|\gamma|\exp(i\theta)$ and the parameter $\phi$ is a controllable phase introduced between the two  optical beams with orthogonal polarizations. As the phase $\phi$ is varied, the intensity  $I(\phi)$ varies accordingly, reaching maxima $I_{max}$ and minima $I_{min}$ of intensity. The visibility is 
\begin{equation}
\label{eq: v}
\text{V}=\frac{I_{max}- I_{min}}{I_{max}+I_{min}}=\frac{2\sqrt{I_H I_V} }{I_H+I_V}|\gamma|.    
\end{equation}
When $I_H=I_V$ the visibility gives the overlap, i.e. $V=|\gamma|$ that is the degree of polarization. When the intensities of the beams with orthogonal polarizations are different, we can make use of the concept of distinguishability~\cite{eberly2017} 
\begin{equation}
\label{eq: d}
D=\frac{|I_H -I_V|}{I_H+I_V}.
\end{equation}
The distinguishability is zero when the two beams have the same intensity, and it reach a value of $1$ when $I_H \gg I_v$ or $I_V \gg I_H$. Using Eq. (\ref{eq: v}) for the visibility and Eq. (\ref{eq: d}) for the distinguishability, we can write that the modulus of the overlap ($|\gamma|$)  is 
\begin{equation}
|\gamma|=\frac{\text{V}}{\sqrt{1-\text{D}^2}}. 
\end{equation}
By measuring the distinguishability $D$ and the visibility $V$ we obtain the overlap $|\gamma|$ that quantifies the degree of similarity between the input spatial shape and the spatial shapes contained in the database. Notice that we also determine the degree of polarization of the optical beam, that is
\begin{equation}
\label{eq: d2}
P=\Big[ 1-(1-D^2)\, (1-|\gamma|^2) \Big]^{1/2}.
\end{equation}
The visibility and distinguishability can be easily measured.

\section{Experimental demonstration of the protocol}

\subsection{Experimental setup}
Figure~\ref{fig: ExperimentalSetup} shows the experimental setup used to quantify, by measuring the degree of polarization, the degree of similarity between two spatial shapes. We use a Gaussian beam generated with a HeNe laser ($\lambda=633~\text{nm}$). The beam is given a diagonal polarization with the help of a polarizer and a half wave plate ($\text{HWP}_1$). It is expanded and collimated using a telescope (lenses $\text{L}_1$ and $\text{L}_2$). With the help of a polarization beam splitter ($\text{PBS}_1$), a mirror and $\text{HWP}_2$, the beam is split into two parallel propagating beams with vertical polarization. Each beam impinges on half of the screen of a spatial light modulator (SLM, Hamamatsu X10768-01, 792 $\times$ 600 pixels with a pixel pitch of 20 µm) prearranged with a hologram. The two modulated beams reflected from the SLM are recombined with the help of $\text{PBS}_2$ and two HWPs ($\text{HWP}_3$ and $\text{HWP}_4$). The angles of  $\text{HWP}_3$ and $\text{HWP}_4$ tune the values of the intensities between the two orthogonally polarized beams. The resulting beam contains in one polarization the input spatial shape and in the orthogonal polarization one spatial shape of the database. The beam leaves $\text{PBS}_2$ through one output port and the other output port is blocked and neglected. The beam is sent through a telescope ($\text{L}_3$ and $\text{L}_4$) for collimation.

\begin{widetext}
\begin{minipage}{0.9\linewidth}
\begin{figure}[H]
\centering
\includegraphics[scale=0.25]{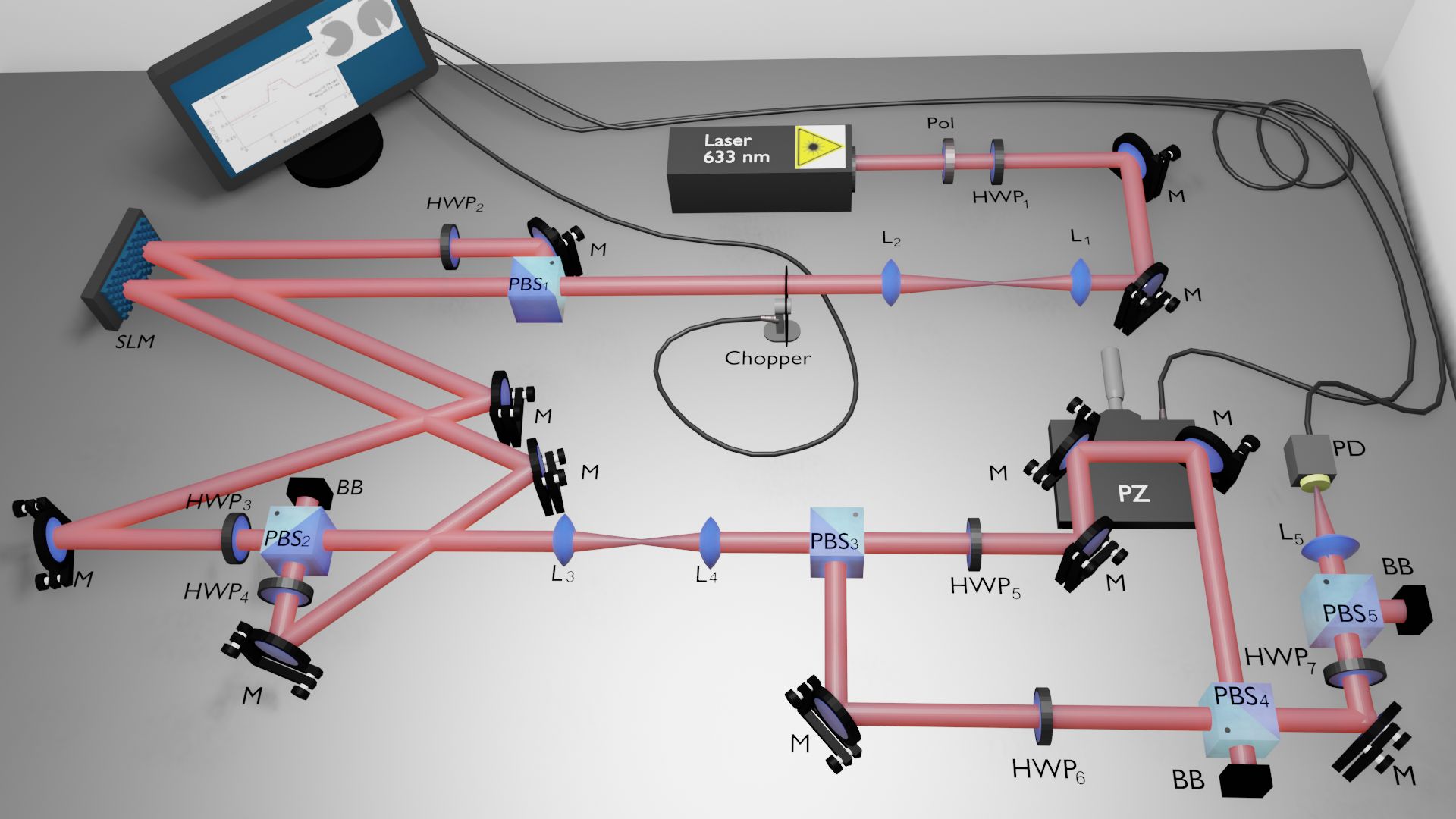}
\caption{ Experimental set-up. PBS: Polarization beam splitter; M: mirrors; HWP: half-wave plates; PZ: piezo-electric; L: lenses; Pol: Polarizer; PD: Photodetector: BB: beam blocker.}
\label{fig: ExperimentalSetup}
\end{figure}
\end{minipage}
\end{widetext}

In the detection stage, the beam traverse a polarization-sensitive Mach-Zehnder interferometer (MZI), where beams with orthogonal polarizations acquire a phase difference by varying the length of one arm of the interferometer with a linear translation stage with piezo-electronics. $\text{HWP}_5$ and $\text{HWP}_6$ flip the polarization to its orthogonal. The output beam of the MZI passes through $\text{HWP}_7$ set at $22.5^\circ$ and $\text{PBS}_7$ to project into diagonal polarization. Light transmitted by  $\text{PBS}_5$ is focused into a photodiode, while light coming from the reflecting output of the $\text{PBS}_5$ is neglected.

The data acquired from the photodiode is processed by a lock-in amplifier system which has as reference the signal of an optical chopper (600 Hz) placed before $\text{PBS}_1$. The integration time of the lock-in amplifier is set at 30~$\text{ms}$. A data acquisition system (DAQ6008) logs 1000 samples of the output signal from the lock-in amplifier while the piezo is moving. 

\subsection{Results and Discussions} 
As first experiment, to show a first proof of the feasibility of the protocol proposed, we generate using the two halves of the SLM, two optical beams:
\begin{widetext}
\begin{eqnarray}
& & \vec{E}_1(\rho,\varphi)=U(\rho)\,\exp(i \varphi) \hat{H}, \label{oam1}\\
& & \vec{E}_2(\rho,\varphi)=U(\rho)\, \Big[ \cos(\theta/2)\,\,\exp(i \varphi)+\sin(\theta/2)\,\exp{-i\varphi} \Big] \hat{V},
\label{oam2}
\end{eqnarray}
\end{widetext}

where $(\rho,\varphi)$ are cylindrical coordinates. $U(\rho)\,\exp(i \varphi)$ corresponds to an optical beam with orbital angular momentum (OAM) per photon of $\hbar \omega$. $U(\rho)$ is the resulting radial profile when the phase imprinted in the SLM is of the form $\text{Mod} \big(2\pi/\Delta + tan^{-1} (y/x),\pi\big)$ \cite{janicijevic2008}, where $\Delta$ is the period of the grating that allows us to separate the diffraction orders. We consider only the first diffraction order. The parameter  $\theta$ is an angle to control the amplitude of the two modes with OAM per photon of $+\hbar \omega$ and $-\hbar\omega$ embedded with vertical polarization. It can be easily shown that the overlap is $|\gamma|=|\cos(\theta/2)|$. 

\begin{figure}[t!]
\centering
\includegraphics[scale=0.35]{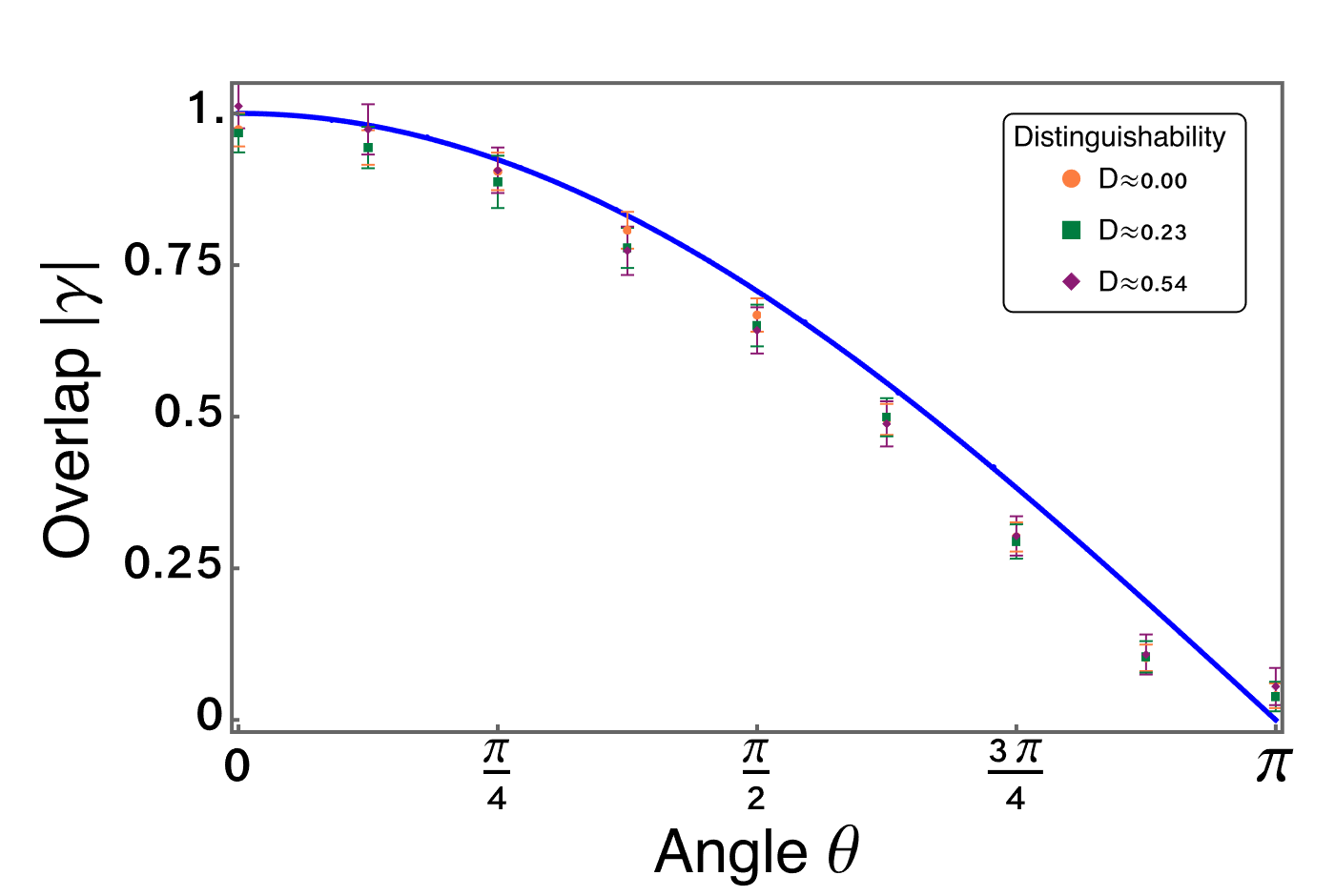}
\caption{Overlap for three different values of the distinguishability for the spatial shapes shown in Eqs. (\ref{oam1}) and (\ref{oam2}). The solid line is the theoretical prediction and the dots are experimental results. The error bars indicate the uncertainty due to propagation of errors in the measurement. }
\label{fig: Experimentaldegree of similarity}
\end{figure}

Figure~\ref{fig: Experimentaldegree of similarity} shows the overlap $|\gamma|$ as a function of the parameter $\theta$. The solid line is the theoretical prediction and the dots correspond the experimental results. For each point the overlap is obtained measuring the visibility and the distinguishability. The visibility is obtained by measuring the interference pattern when the phase difference varies. The distinguishability is obtained by measuring the intensity of each arm of the interferometer. The circle (orange), square (green), and diamond (purple) points are the experimental outcomes when the degree of distinguishability takes values of $D \approx 0$, $D \approx 0.23$, and $D \approx 0.54$, respectively. When $\theta=0$ the two spatial profiles are equal, so the overlap gets its maximum value of $1$. When $\theta$ increases, the overlap decreases until reaching a value of $0$ for $\theta = \pi$. These results show that the protocol to determine $|\gamma|$ is insensitive to differences in intensity between the two optical beams embedded with different spatial beam profiles.

\begin{figure}[t!]
\centering
\includegraphics[scale=0.5]{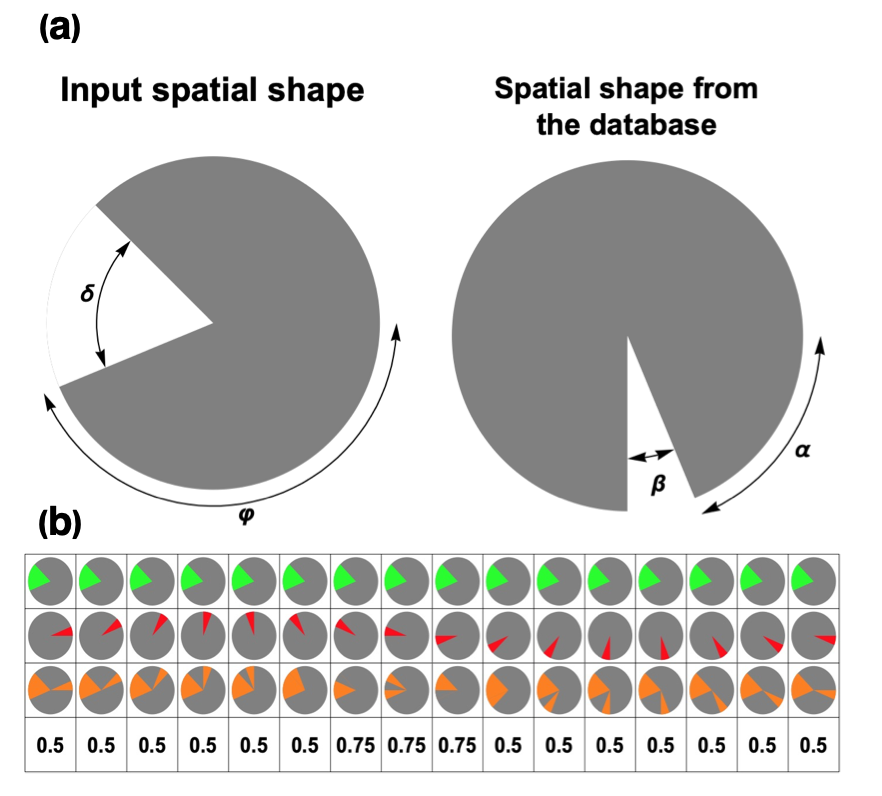}
\caption{(a) Input spatial shape and an example of a spatial shape contained in the database. (b) As example, the input spatial shape is a disk with opening angle $\beta=\frac{3\pi}{8}$ oriented a certain angle as shown in the first row. The second row corresponds to a spatial shape of the database. The opening angle is $\beta=\pi/8$ and is rotated in steps of $\Delta \alpha=\frac{\pi}{8}$ in the interval $\{0, 2\pi\}$. The third row shows with different colors the areas where both spatial shapes coincide (the same value $+1$ or $-1$) and where they do not coincide. The last row shows the value calculated of the overlap.}
\label{fig: DiskExample}
\end{figure}

For the demonstration of the protocol described above, we perform a second set of experiments. We consider the objects shown in Fig. \ref{fig: DiskExample}(a), that are similar, although not identical, to the objects used by Xie et al. \cite{Xie2017} to demonstrate parameter sensing making use of the complex orbital angular momentum spectrum of light beams. The objects are emulated with the help of a spatial light modulator (SLM), that is divided in two halves to allow for the presence of two different spatial shapes. The input spatial shape is generated by a disk with an opening angle $\delta$ oriented an angle $\varphi$ measured from the axis $y=0$. The spatial shapes included in the database are generated with disks with opening angle $\beta$ oriented an angle $\alpha$. In the area of the SLM that corresponds to the opening angle, the pixels of the SLM introduce a zero phase. Everywhere else the phase introduced by the pixels of the SLM is $\pi$, so we can assume that the result of illumination of the object it to multiply the incoming illumination beam by $+1$ or $-1$ depending if the corresponding position of the pixel is inside the opening angle or not. Therefore that overlap is
\begin{equation}
    |\gamma|=\Big| \frac{A_+-A_-}{A_++A_-} \Big|,
\end{equation}
where $A_+$ is the area where both beams have the same phase, and $A_-$ is the area where they have different phases. Since the total area is  $A_T=A_++A_-$, we have
\begin{equation}
|\gamma|=\Big|1- 2\frac{A_{-}}{A_{T}} \Big|.
\label{prediction}
\end{equation}
Fig.~\ref{fig: DiskExample}(b) shows an example of the procedure used to determine the overlap. The first row shows the  input spatial shape that is a disk with opening angle $3\pi/8$. The second row shows a spatial shape of the database with opening angle $\pi/8$ that is rotated in steps of $\Delta \alpha=\pi/8$. The areas with the same or different value are shown in the third row. In the example, $A_{-}/A_{T}$ oscillates between a maximum value of $A_{-}/A_{T}=0.5$ and a minimum value of $A_{-}/A_{T}=0.25$. The fourth row shows the value of the overlap for each step of the scan.

\begin{figure}[H]
     \centering
     \begin{tabular}{@{}c@{}}
         \centering
         \includegraphics[scale=0.3]{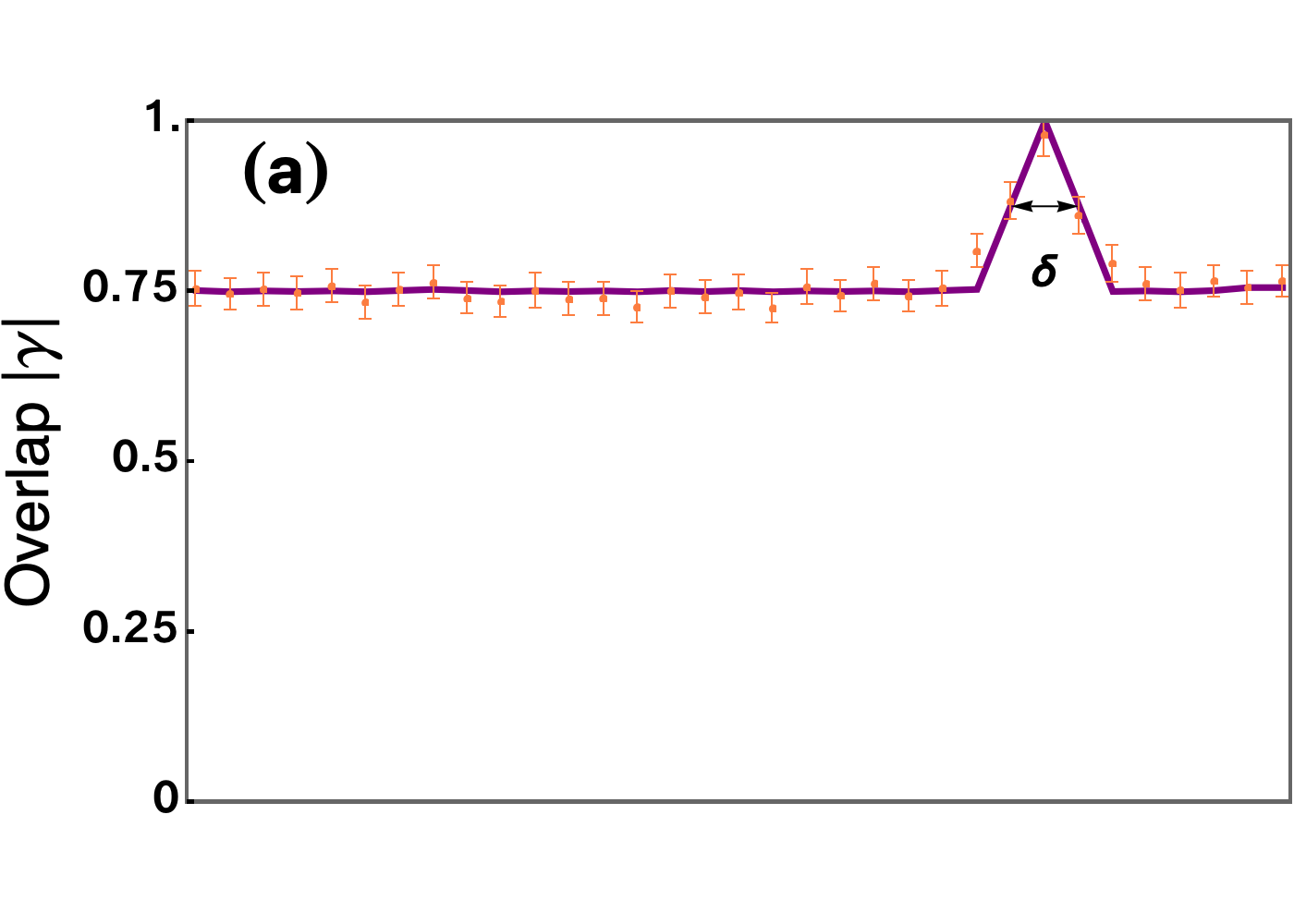}
     \end{tabular}
     \begin{tabular}{@{}c@{}}
         \centering
         \includegraphics[scale=0.3]{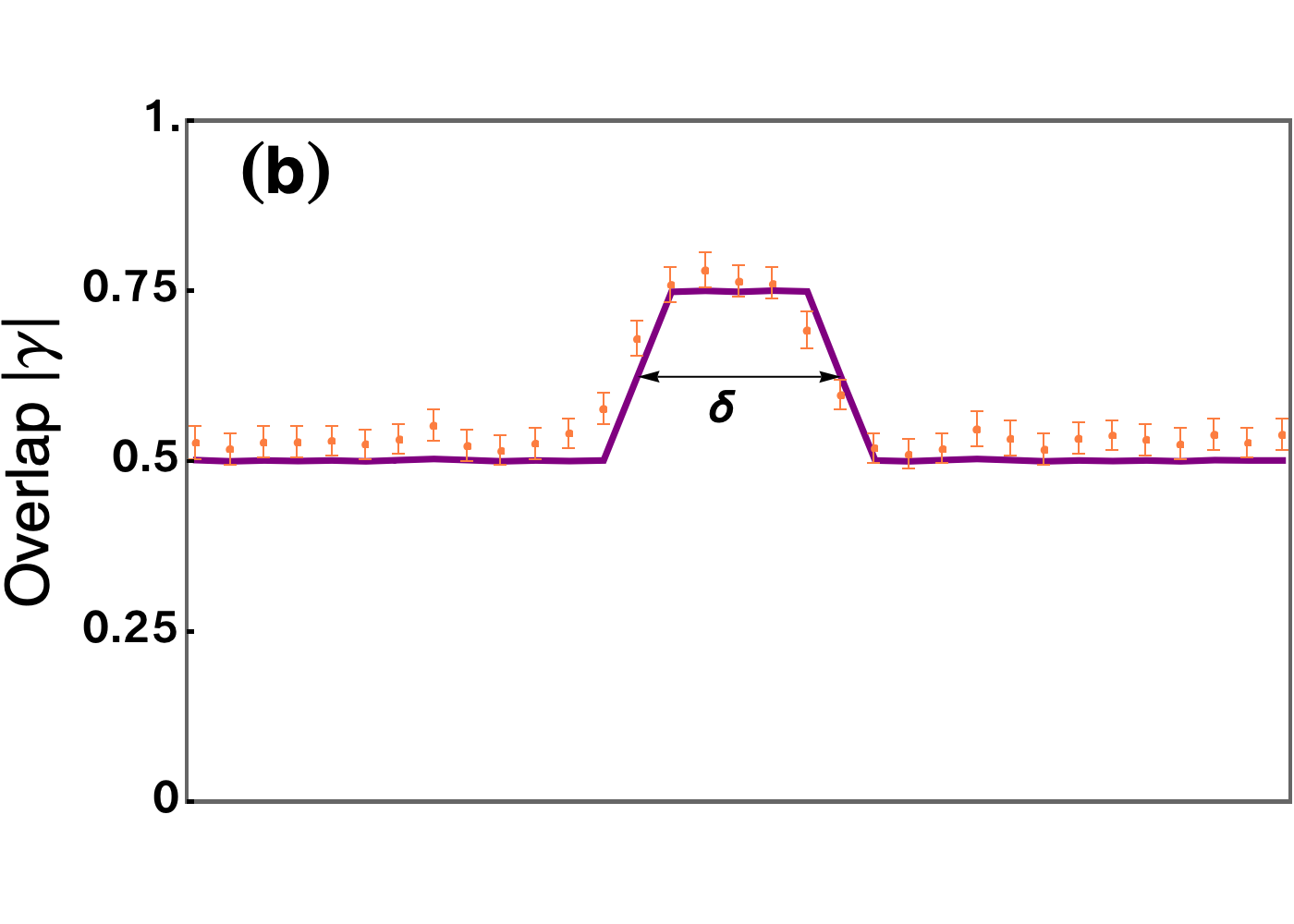}
     \end{tabular}
     \begin{tabular}{@{}c@{}}
         \centering
         \includegraphics[scale=0.3]{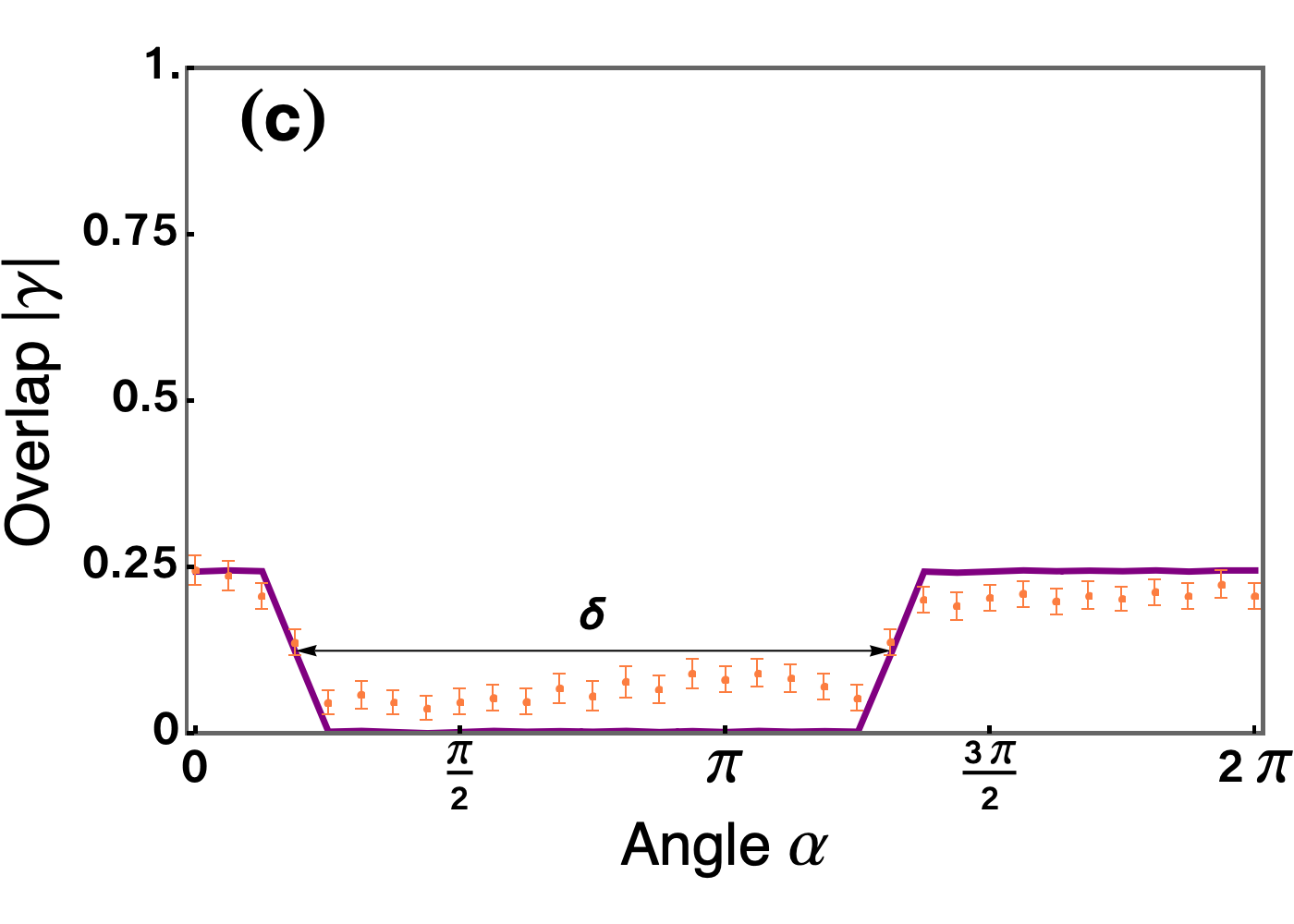}
     \end{tabular}
        \caption{Overlap as function of the angle $\alpha$. (a) Input spatial shape with $\delta=\frac{\pi}{8}$ and $\varphi=\frac{25\pi}{16}$, (b)  Input spatial shape with $\delta=\frac{3\pi}{8}$ and $\varphi=\frac{7\pi}{8}$; (c) Input spatial shape with $\delta=\frac{9\pi}{8}$ and $\varphi=\frac{\pi}{4}$. The spatial shape coming from the database has an opening angle of $\beta=\frac{\pi}{8}$ in all cases. We vary the angle $\alpha$ in steps of $\Delta \alpha=\frac{\pi}{16}$ in the interval $\{0, 2\pi\}$}
        \label{fig: Diskresult}
\end{figure}

Fig.~\ref{fig: Diskresult} shows the experimental results obtained for three different input spatial shapes. The solid line is the theoretical prediction using Eq. (\ref{prediction}), and the dots are the data obtained experimentally. The error bars indicate the uncertainty of the measurement. Fig.~\ref{fig: Diskresult}(a) shows the case when $\delta=\frac{\pi}{8}$ and $\phi=\frac{25\pi}{16}$. Fig.~\ref{fig: Diskresult}(b) shows the case when $\delta=\frac{3\pi}{8}$ and $\phi=\frac{7\pi}{8}$. Finally, Fig.~\ref{fig: Diskresult}(c) shows the case when  $\delta=\frac{9\pi}{8}$ and $\phi=\frac{\pi}{4}$. 

It can be seen that the overlap is $|\gamma|=1$ when the two spatial shapes are equal and have the same orientation, as shown in Fig.~\ref{fig: Diskresult}(a). Otherwise, the overlap never reach the maximum value of $1$. In all cases one observes a similar pattern as function of $\alpha$. There is a  rising and falling edge varying the angle $\alpha$. It can be shown that the distance between the midpoints of the rising and falling edges is the value of $\delta$ (this is shown for the sake of clarity in Fig.~\ref{fig: Diskresult}). The experimental and theoretical values of $\delta$ agree well for the cases represented in Figs.~\ref{fig: Diskresult}(a) and~\ref{fig: Diskresult}(c). For the case represented in Fig.~\ref{fig: Diskresult}(b) there is a slight discrepancy. We obtain an experimental estimation of $\delta$ of $5\pi/16$, when the theoretical value is $3\pi/8$. This might due to the fact that we are varying the angle $\alpha$ in steps of $\pi/16$, that is precisely the discrepancy we observe in this case.

\section{Conclusions}
We have put forward and demonstrated experimentally a protocol to quantify the degree of similarity between two spatial shapes embedded into two orthogonally polarized optical beams. The protocol avoids measuring the amplitude and phase at each point of of both beams, a measurement that is generally cumbersome to do. Instead it measures the degree of polarization of the combined optical beam, a measurement that is generally much easier to do. The measurement of the degree of polarization only requires measuring the power imbalance between two arms of a Mach-Zehnder interferometer and the visibility of intensity fringes as function of the path difference. The measurement is insensitive to a global phase difference between the spatial shapes and to intensity differences between them.

The key enabling tool of the protocol is the use of non-separable beams, where different degrees of freedom can not be considered independently. In our case the two degrees of freedom considered are polarization and spatial shape. The non-separability is also an important feature of quantum entanglement, where the quantum state of entangled separate entities can not be separated in quantum states corresponding to each entity. 

The practical implementation of the method considered for pattern recognition in large databases should require the consideration of technical hurdles such as the need to implement accurately and fast enough a large number of spatial shapes, given the current limitations of the switching speed of spatial light modulators. Another important issue might be how the presence of speckle can influence the accuracy of the estimation of the degree of similarity between two spatial shapes.

The protocol presented here is thus another example of a quantum-inspired technology. The inspiration comes from the observation that measuring the purity (first-order coherence) of a single photon of a bipartite two-photon state is a quantifier of the overlap between the two states of the companion photon with whom the single photon measured can be correlated. This specific feature of quantum entangled states can be mimicked classically making use of non-separable optical beams. This is a first step toward the implementation of a new kind of {\it image comparator}, an image comparison analysis tool where images are compared without the need to measure and analyze its shapes. This is an interesting practical consequence of the theory of coherence of non-separable and non-uniform optical beams.  

The implementation of quantum-inspired classical protocols can provide algorithms with similar performance to the corresponding quantum protocols, but that might be much easier to implement. One example is Grover's search algorithm, a quantum-mechanical technique for searching N possibilities in only $\sqrt{N}$ steps. Grover and Sengupta \cite{Grover2002} found that a similar
algorithm applies in a purely classical setting, and that it still shows advantageous performance when compared with alternative methods.

\section*{Acknowledgements}
This work is part of the R$\&$D project CEX2019-000910-S, funded by the Ministry of Science and innovation (MCIN/ AEI/10.13039/501100011033/), from Fundació Cellex, Fundació Mir-Puig, and from Generalitat de Catalunya through the CERCA program. We acknowledge suport from the project 20FUN02 “POLight”, that has received funding from the EMPIR programme co-financed  by the  Participating  States  and  from  the  European  Union's  Horizon  2020  research  and innovation programme. JPT also acknowledges financial support from project QUISPAMOL (PID2020-112670GB-I00) funded by MCIN/AEI /10.13039/501100011033. DFU acknowledge financial support (PRE2018-085072) financed by   MCIN/AEI/10.13039/501100011033 and FSE ``El FSE invierte en tu futuro''

\bibliography{References}

\end{document}